# Concurrency Control for Adaptive Indexing

Goetz Graefe[†]    Felix Halim[‡]    Stratos Idreos[⋆]    Harumi Kuno[†]    Stefan Manegold[⋆]

[†]HP Labs, Palo Alto, CA, USA    [‡]National University of Singapore    [⋆]CWI Amsterdam, The Netherlands
{harumi.kuno, goetz.graefe}@hp.com    halim@comp.nus.edu.sg    {idreos, manegold}@cwi.nl

## ABSTRACT

Adaptive indexing initializes and optimizes indexes incrementally, as a side effect of query processing. The goal is to achieve the benefits of indexes while hiding or minimizing the costs of index creation. However, index-optimizing side effects seem to turn read-only queries into update transactions that might, for example, create lock contention.

This paper studies concurrency control in the context of adaptive indexing. We show that the design and implementation of adaptive indexing rigorously separates index structures from index *contents*; this relaxes the constraints and requirements during adaptive indexing compared to those of traditional index updates. Our design adapts to the fact that an adaptive index is refined continuously, and exploits any concurrency opportunities in a dynamic way.

A detailed experimental analysis demonstrates that (a) adaptive indexing maintains its adaptive properties even when running concurrent queries, (b) adaptive indexing can exploit the opportunity for parallelism due to concurrent queries, (c) the number of concurrency conflicts and any concurrency administration overheads follow an adaptive behavior, decreasing as the workload evolves and adapting to the workload needs.

## 1. INTRODUCTION

This paper focuses on concurrency control for read-only queries in adaptive indexing. Adaptive indexing enables incremental index creation and optimization as automatic side effects of query execution. The adaptive mechanisms ensure that only tables, columns, and key ranges with actual query predicates are optimized [23, 11, 12, 14, 22, 20, 21]. The more often a key range is queried, the more its representation is optimized; conversely, columns that are not queried are not indexed and indexes are not optimized in key ranges that are not queried. Prior research has introduced adaptive indexing in the forms of database cracking [20, 21, 22, 16], adaptive merging [14, 12] as well as on hybrids [23] and benchmarking [11]. Past work focused on algorithms and data structures as well as on the benefits of adaptive indexing over more traditional index tuning and on workload robustness.

**The Problem: Read Queries Become Write Queries.** In adaptive indexing, queries executing scans or index lookups may invoke operations that incrementally refine the database's physical design as side effects of query execution. Refinement operations construct and optimize index structures, causing logically "read-only" queries to update the database. This raises the question whether the concurrency control required to support these updates incurs serious overhead and contention.

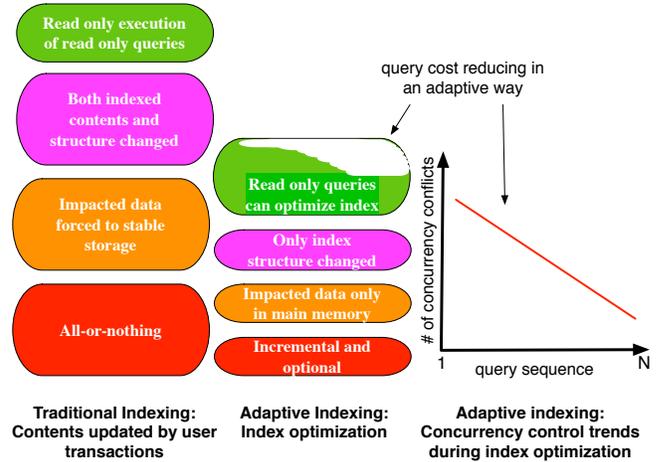

**Figure 1: Adaptive versus explicit indexing.**

**Index Contents vs. Index Structure.** With regard to the concurrency control required to coordinate index updates, refining and optimizing an adaptive index during read queries is much simpler than updating a traditional index. Figure 1 illustrates the underlying intuition, comparing the incremental refinement of adaptive indexing to the explicit index updates involved in traditional indexing. The heights of each pair of boxes roughly illustrate the relative costs of various characteristics. Unlike traditional systems, in adaptive indexing, execution of read-only queries can trigger index updates and improves adaptively over time. On the other hand, the index changes caused by read queries impact only physical index structures, never logical index contents, and thus (a) concurrency can be governed using only short-term in-memory latches as opposed to transactional locks and (b) the purely structural updates are optional and can be *skipped or pursued opportunistically*. These distinctions relax constraints and requirements with regard to concurrency control of adaptive indexing compared to those of traditional explicit index updates and enable new techniques for reducing the performance overhead of concurrency control during structural index updates.

**Incremental Granularity of Locking.** Another powerful characteristic of adaptive indexing is that the more an index is refined, the better index structures support concurrent execution by enabling a finer granularity of locking. That is to say, refinements to an index's structure enables updates to acquire increasingly precise locks. This effect is shown in the right part of Figure 1, which illustrates the number of conflicts decreasing as the workload sequence evolves. Thus, as in query processing, concurrency control for adaptive indexing dynamically adapts to the running workload.

**Contributions.** The current paper explores and proposes techniques for concurrency control that reduce the overhead imposed by adaptive indexing on read-only query execution to negligible levels. More specifically, we show the following:

- Adaptive indexing maintains its adaptive properties during the execution of concurrent queries.

- Concurrency conflicts adaptively decrease as adaptive indexing adjusts to the running workload.

- Adaptive indexing can exploit concurrent queries to increase parallelism.

In this paper, we focus on logically "read-only" queries that update the index only as a side effect of processing. We note that update algorithms for adaptive indexing have already been studied in [22] and that read-write conflicts in concurrent access can be resolved with the techniques reported here with minor modifications.

**Paper Organization.** Section 2 places this effort into the context of prior work. Section 3 discusses how adaptive indexes perform purely structural and opportunistic changes, relaxing constraints and requirements and thus enabling new techniques such as incremental locking and adaptive early termination. Sections 4 and 5 look at two concrete examples of how these new techniques can be applied using different adaptive indexing methods. Then, Section 6 presents a detailed experimental analysis over a column-store system, demonstrating that any concurrency control overheads are minimal and that they adapt to the workload. Finally, Section 7 previews our ongoing work and Section 8 presents a summary and discusses conclusions.

## 2. PRIOR WORK

Many prior index tuning and management approaches focus on optimizing decisions related to the management of full index structures that cover all key ranges [3, 4, 5, 17, 19, 2, 29]. Some recognize that some data items are more heavily queried than others and support partial indexes [27, 31], while others recognize that not all decisions about index selection can be taken up front and provide online indexing features [2, 29]. In either case, explicitly creating and refining index structures using independent operations, as opposed to as a side effect of query processing, does not impose additional concurrency overhead upon the processing of read-only queries; full or partial indexes are created either up front or periodically, interleaving query execution.

The defining characteristic of adaptive indexing is that indexes are created and refined *incrementally and continuously* as a side effect of query processing. This brings automatic adaptation of the physical design to the workload, but also introduces concurrency control issues during (adaptive) indexing. Although recent surveys on concurrency control and recovery [8, 10] cover these topics for B-tree indexes and have influenced our research, to our knowledge the present paper is the first to focus explicitly on how adaptive indexing mechanisms can support the transactional guarantees of the underlying database management system, without imposing undue overhead on the processing of read-only queries.

Below, we describe a variety of adaptive indexing, or adaptive-indexing-like, mechanisms.

**Database Cracking.** "Database cracking" pioneered focused, incremental, automatic optimization of the representation of a data collection — the more frequently a key range is queried, the more its representation is optimized for future queries [20, 21, 22, 16, 24]. As its name suggests, database cracking splits an array of values into increasingly refined partitions. One can think of it as an incremental quicksort where each query may result in a partitioning step. Index optimization is entirely automatic and occurs as a side effect of queries over key ranges not yet fully optimized.

For example, Figure 2 shows data being loaded directly, without sorting, into an unsorted array. A read-only query on the range "d – i" then arrives. As a side effect of answering that query, the array is split into three partitions: (1) keys before 'd'; (2) keys that fall between 'd' and 'i'; and (3) keys after 'i'. Then a new query with range boundaries 'f' and 'm' is processed. The values in partition (1) can be ignored, but partitions (2) and (3) are further cracked on keys 'f' and 'm', respectively. Subsequent queries continue to partition these key ranges until the structures have been optimized for the current workload.

**Adaptive Merging.** Inspired by database cracking, "adaptive merging" also refines index structures during query processing [12, 14]. While database cracking resembles an incremental quicksort, with each query resulting in at most two partitioning steps, adaptive merging resembles an incremental external merge sort. In adaptive merging, the first query with a predicate against a given column produces sorted runs. Each subsequent query against that column then applies at most one additional merge step to each record in the desired key range. All records in other key ranges are left in their initial or current places. As with database cracking, this merge logic takes place as a side effect of query execution.

For example, Figure 3 shows an initial read-only query that creates equally-sized partitions and sorts them in memory to produce four sorted runs. While a second query with range boundaries 'd' and 'i' is processed, relevant values would be retrieved (via index lookup because the runs are sorted) and merged out of the runs and into a "final" partition. Similarly, results from a third query with range boundaries 'f' and 'm' are merged out of the runs and into the final partition. Subsequent queries continue to merge results from the runs until the "final" partition has been fully optimized for the current workload.

**Hybrid Adaptive Indexing.** Database cracking and adaptive merging have distinct strengths; our adaptive indexing "hybrid" approach brings together both sets of strengths in the context of an in-memory column store [23]. Each step of database cracking is like a single step in a quicksort, whereas the first step of adaptive merging creates runs, which subsequent steps merge. Thus, database cracking enjoys a low initialization cost, but converges relatively slowly, whereas adaptive merging has a relatively high initialization cost but converges quickly to an optimally-refined index.

Our hybrid adaptive indexing algorithms apply different refinement strategies to initial versus final partitions, exploiting the insight that in adaptive merging, once a given range of data has moved out of initial partitions and into final partitions, the initial partitions will never be accessed again for data in that range. A final partition, on the other hand, is searched by every query, either because it contains the results or else because results are moved into it. Therefore, effort that refines an initial partition is much less likely to "pay off" than the same effort invested in refining a final partition.

The hybrid algorithms combine the advantages of adaptive merging and database cracking, while avoiding their disadvantages: *fast convergence, but without the burden of fully-sorting the initial partitions.* For example, Figure 4 shows an initial read-only query that creates four equally-sized unsorted initial partitions. While a second query with range boundaries 'd' and 'i' is processed, each initial partition is cracked on the query's boundaries, and the requested values are merged out of the initial partitions and into a sorted "final" partition. Similarly, from a third query with range boundaries 'f' and 'm', the initial partitions that hold relevant values are cracked, and the result values are merged out of the initial partitions and into the final partition. Subsequent queries continue to crack initial partitions and merge results from them until the "final" partition has been fully optimized for the current workload.

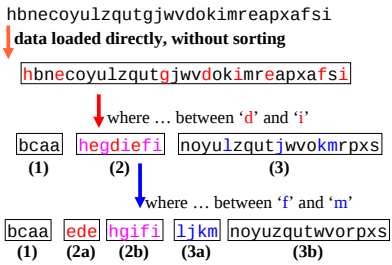

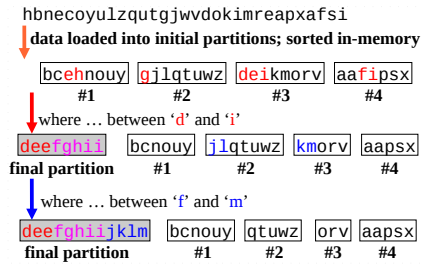

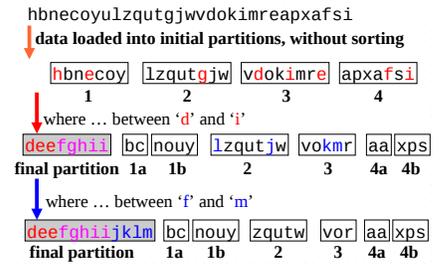

Figure 2: Database cracking.

Figure 3: Adaptive merging.

Figure 4: Hybrid "crack-sort".

**Soft Indexes.** "Soft indexes" automatically and autonomously manage indexes in response to a workload [25]. Like the monitor-and-tune approaches, this approach continually collects statistics for recommended indexes and periodically and repeatedly automatically solves the index selection problem. Unlike typical monitor-and-tune approaches and like adaptive indexing approaches, [25] then generates (or drops) the recommended indexes as a part of query processing. Unlike adaptive indexing approaches like database cracking and adaptive merging, however, neither index recommendation nor creation is incremental; explicit statistics are kept and each recommended index is created and optimized to completion (although the command might be deferred). Although we recognize soft indexes as kin to database cracking and adaptive merging, in the remainder of this paper we focus upon the latter, i.e., incremental and adaptive indexing methods.

## 3. APPROACH

While our prior work on adaptive indexing [11, 12, 14, 20, 21, 22, 16, 23, 24] has focused on data structures and algorithms, transactional guarantees are also required for integration of new techniques into a database management system. We are informed by recent surveys on concurrency control and recovery [8, 10] that cover these topics for B-tree indexes. The transactional ACID guarantees include (failure) atomicity, consistency, isolation (synchronization atomicity), and durability. Database systems usually implement recovery (failure atomicity, durability) by write-ahead logging and concurrency control (synchronization atomicity) by locking and latching.

Adaptive indexing introduces new potential states to the index life cycle. The diagram in Figure 5 compares the relationship between index states in traditional online index operations versus in adaptive indexing. State 2 is invisible and of very short duration in most systems. An exception is Microsoft SQL Server, where this state is called a "disabled index". In State 3, the index is partially populated, i.e., it contains fewer index entries than there are rows in the underlying table, but the index is fully optimized. That is, those key ranges already in the index are in their final position within the index. Table and index can be updated (that's the "online" aspect of traditional online index creation) but the index cannot be used for search during query processing.

Adaptive indexing refinements take place in State 4, where the index is fully populated but not fully optimized. In other words, all

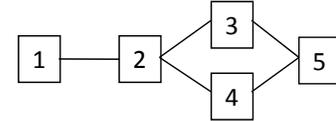

**States**
1. Index does not exist
2. Index is in the catalogs but does not contain any records
3. Partial index in a traditional online index operation
4. Partially optimized index in adaptive indexing
5. Finished index – fully populated and fully optimized

Figure 5: Traditional online vs. adaptive indexing.

index entries exist but not yet in their final position. In this state, the fully populated, partially optimized index is available for both read-only query processing and read-write update processing, whereas the partially populated, fully optimized partial index in online index creation requires effort in all updates but does not contribute to read-only query processing. Optimization of the index is left to future query execution and will affect only those index entries relevant to actual queries, i.e., key ranges in actual predicates. Optimization of other key ranges is deferred until relevant queries are encountered, possibly indefinitely. A single user transaction might encounter multiple such cases, e.g., querying and updating multiple key values in a single index which is being optimized by adaptive indexing techniques.

Fundamentally, two factors mitigate the concurrency control overhead that adaptive indexing incurs when executing queries that are logically "read-only" but which refine index data structures. First, with read-only queries, adaptive indexing performs only *structural* modifications to the physical representation of the index, leaving the logical *contents* of the index unmodified. This separation between user data and system state is very powerful and gives the system transactions of adaptive indexing independence from user transactions, even if they run within the same execution thread. For example, if the user transaction rolls back for some reason, there is no need to reverse the index optimization already achieved. More subtly, the user transaction (and its query) might run in a low transaction isolation level, e.g., read committed, whereas the index optimization must achieve complete correctness and synchronization atomicity with respect to all other transactions active in the system.

Second, the adaptation of index data structures to conform to the current workload enables the automatic and dynamic adaptation of the lock granularity of locks needed to coordinate structural changes. That is, as the workload progresses and the physical data structures become increasingly refined, not only do structural changes become less likely, but also the objects locked by refinement operations become increasingly finer-grained, reducing the likelihood of contention.

### 3.1 Locks vs. Latches

The usual understanding of physical data independence focuses on tables and indexes. In addition, some data structures such as B-trees (and their variants) permit multiple representations for the same logical index contents. For example, a B-tree node may be compressed (shortened records) or compacted (no free space fragmentation), it may contain "pseudo-deleted" "ghost" records (left by a deletion), etc. Similarly, boundary keys between nodes might be chosen by record count or by byte count, by length of the separator key [1], by desired "fill factor" (e.g., 90% during database loading), etc.

Table 1: Locks and latches.

|  | index *Locks* | *Latches* |
| --- | --- | --- |
| Separate... | User transactions | Threads |
| Protect... | Database contents | In-memory data structures |
| During... | Entire transactions | critical sections |
| Modes... | Shared, exclusive update, intention, escrow, schema, etc. | Reads, writes, (perhaps) update |
| Deadlock... | Detection & resolution | Avoidance |
| ...by... | Analysis of the waits-for graph, timeout, transaction abort, partial rollback, lock de-escalation | Coding discipline, "lock leveling" |
| Kept in... | Lock manager's hash table | Protected data structure |

The separation between logical index contents and physical data structure or representation affects the mechanisms used to enact their concurrency control. Locks separate transactions and protect logical contents, including the empty gaps between existing keys in serializable key range locking, whereas latches separate threads and protect data structures present in memory. Table 1, taken from [8], summarizes their differences. The crucial enabler is the separation of logical contents and physical structure.

## 3.2 Hierarchical and Incremental Locking

In traditional systems, the granularity of locking is fixed over time — multiple granularity choices may be available, but once chosen, remains fixed for the system as a whole. For example, a workload made up of a multitude of concurrent small transactions might use fine-grained locking of individual keys, whereas coarser locks would enable large transactions to lock large ranges of keys efficiently without having to acquire a multitude of locks.

In order to reduce the number of locks required, hierarchical locking within an index can be employed. Hierarchical locking is a special case of multi-granularity locking that enables multitudes of small and large transactions to execute concurrently [7]. The key idea of hierarchical locking is that database objects must be locked according to their containment hierarchies. For example, a transaction that wanted to lock a leaf page in a B-tree index would first acquire a read lock on the table or view, then lock the index or index partition, and then finally lock the page. Multiple transactions could thus concurrently lock various leaves, and a subsequent large transaction could easily identify whether it can lock a partition without having to check each individual leaf's status.

Several designs for hierarchical locking in index exist, e.g., key range locking on separator keys within a B-tree index or on key prefixes of various lengths [7]. For example, if the artificial leading key field in a partitioned B-tree is a 4-byte integer, Tandem's "generic lock" applied to the 4-byte prefix effectively locks an entire partition [7, 28].

Hierarchical locking is limited to a pre-defined hierarchy of data structures, e.g., key, page, and index. The key idea of incremental locking is that the lock granularity can be changed dynamically, to adapt to the current workload. For example, given a workload that consists entirely of a multitude of small transactions in the morning and then shifts in the afternoon to eventually consist entirely of key range operations, an incremental locking system would automatically and dynamically shift from locking individual keys to locking key ranges. The partitions created by database cracking are naturally conducive to incremental locking, in that the partitions created as a side effect of index refinement also represent sub-objects that can then be locked by subsequent operations.

## 3.3 Concurrency Control

The following focuses on a single-threaded query with index optimization and on concurrency control with respect to other queries. This scenario is particularly relevant with regard to State 4 as shown in Figure 5, where an index has been created and added to the catalogs, but the index has not yet been fully optimized for this workload, so queries may still result in updates to index structures.

**Concurrency Control by Latching.** Since index optimization affects only index structure, not logical index contents, the thread and system transaction performing the index reorganization may rely entirely on latches. There is no need for acquisition of any locks, although it is required to verify that no concurrent user transaction holds conflicting locks. The latches (on index pages) are retained during a quick burst of reorganization activity; as in standard system designs, user transactions cannot request locks on a page or on key values within an index page without first acquiring the latch on the page.

**Conflict Avoidance.** Index reorganization in adaptive indexing is *optional*. Adaptive indexing treats each read query as an opportunity to improve the physical design. All actions are optional, as adaptive indexing inherently operates on incomplete, not fully optimized indexes. In other words, if an individual query fails to optimize the index, some other query will do so soon thereafter if necessary — and the bigger the potential impact of the refinement action, the more likely that it will eventually take place. Thus, if a query intends to optimize an index but finds that some concurrent user transaction holds conflicting locks, the query can simply forgo the index optimization.

**Early Termination.** Some forms of adaptive indexing can terminate an optimization step at any time yet leave behind a consistent and searchable index, which subsequent queries and their side effects may optimize further. Thus, if a user transaction attempts to access pages latched by an active system transaction performing index optimization, the system transaction may terminate instantly, release its exclusive latches, or downgrade them to shared latches, permitting the concurrent user query to proceed.

**Implicit Multi-version Concurrency Control** Finally, adaptive indexing lends itself naturally to multi-version concurrency control. Because index structures are independent from contents, two transactions may each operate upon their own copy of a contended index structure, which may be assigned version numbers.

## 3.4 Summary

In summary, efficient concurrency control requires strict separation of logical index contents and physical index structure. Reorganization of an index, whether database cracking or adaptive merging, does not affect its logical contents. Thus, index optimization can avoid user transactions and locks. Instead, it can rely on system transactions, latches, and many small transactions with low overheads for invocation and commit processing. These system transactions must respect existing locks held by user transactions but the system transactions have no need to acquire and retain locks.

Moreover, some optimizations are specific to contents-neutral index operations. In case of concurrency contention, a system transaction may simply stop, commit work already completed, and defer further planned work to a subsequent system transaction. We are currently considering whether these properties are more general, e.g., apply to all system transactions, not only index optimization in adaptive indexing techniques.

## 4. ADAPTIVE INDEXING IN B-TREES

The proposed data structure for the adaptive merging is a partitioned B-tree, which is a traditional B-tree index with an artificial leading key field that captures partition identifiers [6]. Therefore, many techniques designed for B-tree indexes can be used, not only with respect to data structures, storage management, etc. but also with respect to concurrency control.

### 4.1 Partitioned B-tree

Partitioned B-trees are an ideal foundation for adaptive indexing as realized by adaptive merging. There are three differences between a partitioned B-tree and a B-tree partitioned in a traditional parallel database management system. First, a partitioned B-tree is a single B-tree, whereas a traditional partitioned index employs a B-tree per partition. Second, partitions in a traditional partitioning scheme are listed in the database catalogs, whereas a partitioned B-tree contains multiple partitions simply by means of distinct values in the artificial leading key field. Third, partitions in a traditional partitioning scheme require catalog updates with the attendant concurrency control protocols, e.g., exclusive locks on data and metadata of a table, whereas partitions in a partitioned B-tree appear and disappear simply by insertion and deletion of records with appropriate values in the artificial leading key field. Each individual B-tree in a traditional partitioning scheme might actually be a partitioned B-tree in order to support efficient index creation and incremental loading.

### 4.2 Transactions and Partitioned B-trees

Transactional guarantees in adaptive merging rely on combining partitioned B-trees with the techniques outlined in Section 3. Partitioned B-trees can capture a wide variety of intermediate states during external merge sort and during B-tree creation, enabling simple and efficient implementations of adaptive merging

As B-trees, they can inherit proven concepts and implementation techniques for concurrency control. One significant advantage of this is that index creation and reorganization don't require logging detailed index contents.

Moreover, adaptive indexing focuses on optional indexes, i.e., those neither created nor prohibited by a tuning tool or a database administrator. Such indexes can be dropped at any time. Re-creation of such an index can exploit knowledge gained during earlier query execution. The side effects of earlier queries may be re-created in the new index even without merging.

Finally, no query relies on a specific earlier query sequence or specific optimization or completion of the B-tree index. Thus, several techniques for efficient concurrency control rely on interpreting any requested index optimization as optional. For even more flexibility, input and output pages in merge steps can enable multi-version concurrency control sufficient to separate read-only queries and queries with index optimization as side effect.

If adaptive indexing is implemented with partitioned B-trees and thus traditional B-tree indexes, concurrency control and recovery can rely on the techniques explored in earlier research and development, e.g., [10, 8, 15, 18]. In the following, we point out specific techniques that enable B-tree-specific adaptive indexing, exemplified by adaptive merging, with low overhead and low contention.

Like other forms of adaptive indexing, adaptive merging relies on a form of differential files [30] for high update rates. In a partitioned B-tree, multiple new partitions might be created during a single load operation. Typically, the size of each new partition is equal to (or twice) the size of the memory available for sorting arriving records. In the context of adaptive indexing, updates and deletions may be applied immediately in place or they may be deferred by insertion of "anti-matter" (deletion markers), which are used routinely in online index creation and in incremental maintenance of materialized views [9]. Insertions may be collected in new partitions within the partitioned B-tree or they may be applied to an existing partition. New partitions seem more appropriate for key ranges not yet optimized (remaining in initial partitions created during index creation), whereas immediate maintenance of an existing partition seems more appropriate for fully optimized key ranges (merged into a single partition during earlier query processing with side effects).

In a partitioned B-tree, each initial run forms its own partition. Similarly, when partitions are merged, the results form a new partition. Partition contents are managed using a table of content data structure. Earlier work has proposed alternative data structures for the table of contents [13].

Many techniques for concurrency control, logging, and recovery have been developed in the context of B-tree indexes [26, 8, 9, 10]. Not surprisingly, most of the new techniques in Section 3 immediately apply to adaptive merging within partitioned B-trees.

### 4.3 Concurrency control

Early research on concurrency control in B-trees failed to separate short-term protection of the data structure versus long-term protection of B-tree contents. The distinctions of contents versus representation, user transactions versus system transactions, locks versus latches, etc. are now standard in sophisticated B-tree implementations. Key range locking for leaf keys is also standard, and key range locking for separator keys explicitly relies on the structure of B-trees. Thus, all these techniques immediately apply to adaptive merging implemented with partitioned B-trees.

A partitioned B-tree is a valid B-tree index, with respect to both contents and representation, independent of the merge steps completed. Records may freely move among partitions. The original partitioned B-trees [6] exploited this property in various ways. It can also be exploited for concurrency control in adaptive merging. In particular, concurrency control conflicts can be avoided or resolved by instantly committing an active merge step and its result.

Finally, merge steps take records from many existing B-tree pages and write new pages in a new B-tree partition. These separate sets of pages readily enable a limited form of multi-version concurrency control, with shared access to the old pages and exclusive access to the new pages until they are committed.

## 5. ADAPTIVE INDEXING FOR COLUMN-ORIENTED DATABASES

In this section, we study the implications of concurrency control for adaptive indexing in a column-store environment. Adaptive indexing was originally proposed as a column-store-specific index mechanism in the form of database cracking [20] and has subsequently evolved to further column-specific refinements such as sideways cracking [22] and hybrid adaptive indexing techniques [23]. Given that the same core principles apply for all adaptive indexing methods, for simplicity of presentation, our discussion fo-

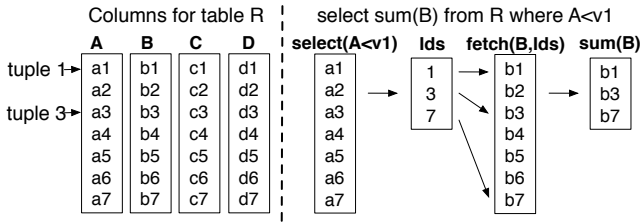

Figure 6: Storage and access in a column-store system.

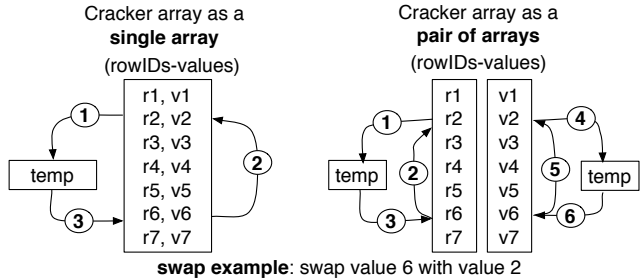

**swap example**: swap value 6 with value 2
Figure 7: Cracker array implementation and swap actions.

cuses mainly on selection cracking [20] (see also last part of Section 5 and Section 7 for more discussion on this).

## 5.1 Column-oriented Storage and Access

The storage and access patterns significantly affect the way concurrency conflicts appear and how they can be resolved. In a column-store system, data is stored one column at a time; every attribute of a table is stored separately as a dense array. This representation is the same both in memory and on disk. All columns of the same table are aligned which allows for efficient positional access to collect all values of a given tuple. For example, all attribute values of tuple $i$ of table $R$ appear in the "$i$-th" position in their respective column. Such an example is shown in the left part of Figure 6.

During query processing, the system accesses one column at a time in a bulk processing mode. The right part of Figure 6 shows the steps of evaluating a simple select-project query in a column-store system. It first evaluates the complete selection over one column. Then, given a set of qualifying IDs (positions), it fetches only the required values from another column before computing the complete aggregation in one go again.

There are two column-store-specific features that adaptive indexing exploits. First, given the underlying representation of data in the form of fixed-length dense vectors, index refinement actions can be implemented very efficiently. Second, due to bulk processing, each column referenced in a query plan is actually used for only a brief period of time compared to the total time needed to process the complete query. For example, as the right part of Figure 6 shows, column $A$ is relevant only for the select operator and is not used for the remainder of the query plan. This means that adaptive indexing only needs to use short-term latches that do not necessarily span the whole duration of a query plan.

## 5.2 Algorithms and Data Structures

In this subsection, we dive deeper into the details of original database cracking [20] to highlight the design issues and data structures that impact concurrency control.

Database cracking relies on continuous but small index refinement actions. Each such action reflects a data reorganization action of the dense array representing the cracking index. In its original design, the cracker index for a column consists of two data structures: (1) a densely populated array of rowID-value pairs that holds an auxiliary copy of the original column of key values, and (2) a memory resident AVL tree that serves as a table-of-contents to keep track of the key ranges that have been requested so far. Each select operator call uses the AVL tree to identify the parts of the index that need to be refined.

The array is continuously physically re-organized (incrementally sorted based on key values) as a side effect of query processing. The nodes in the AVL tree point to the segments ("pieces") in the cracker array where requested key ranges can be found after the respective reorganization step. Thus, the AVL tree provides instant access to previously requested key ranges, and restricts data access as much as possible for the case of a non-exact match, pointing to the shortest possible qualifying range for further cracking.

The latest generation of the cracking release uses a different format for the cracker array. Instead of using an array of rowID-value pairs, it uses a pair of arrays. In the latter case, we have the *rowIDs array* and the *values array*. Figure 7 shows an example comparing both representations. Maintaining separate areas can improve query processing performance e.g., by providing better cache locality for operators that need to access only the rowID array or only the value array.

## 5.3 Concurrency Control

It is sufficient to use rather short-term latches on the cracker array, the AVL tree and some global data structure that keeps track of which cracker indexes do exist.

**Column latches.** For example, consider simple queries that only perform a single selection over a single column; such a query consists of a single select operator that in a bulk mode consumes the entire column and produces the result. When the select operator starts, it first latches the global data structure to check whether a cracker index has already been initialized for the given column. If not, it initializes the respective raw cracker index for that column and latches both the AVL tree and the cracker array. If the cracker index already exists, it latches the AVL tree and the cracker array. Once the latches are acquired, the global data structure can be released, and the select, including any cracker array refinement, is performed with exclusive access to the cracker array. As soon as the select operation, including the necessary array refinement and AVL tree update, finishes, the index latches can be released.

In case of operator-at-a-time bulk-processing as in MonetDB, the select must finish before any other operation in the query plan (that uses the selection result) can start. While using simple coarse-grain per-column latching, this approach benefits from the fact that (1) the latches need to be held only while the select operation is active, and (2) as more queries are processed, both the selection itself and the index refinement benefit from the continuously improving index, shortening the length of the critical section.

**Read-Write Latches.** A more complex scenario is when the same column used for selection (cracking) by one query is also used for aggregation by another query. Reorganizing an array that is being concurrently processed by an aggregation operation that reads every tuple within a qualifying range (e.g., sum or average) may lead to incorrect aggregation results. However, multiple aggregation operations may proceed in parallel over the same column. For this reason, we distinguish between read and write latches. Every cracking select operator requires a write latch over the relevant column; all other, non-cracking, operators require read latches so as to protect the data being read by concurrent cracking operators.

**Example of Column Latches.** The upper two-thirds of Figure 8 illustrates how column latches work for three example queries that arrive concurrently and request access to the same column. For each technique being illustrated, the figure depicts when each query acquires a read (blue dashed line) or write (red solid line) latch. For example, reading the first example ("column latch," top row) from left to right, the three queries arrive concurrently, each requesting to compute a sum over a target range. Thus each query will first crack, and then aggregate over the qualifying range. Initially, all queries request a write latch but only one may proceed, (Q1 in our example). When Q1 finishes with its crack select operator, Q2 wakes up and starts cracking the column for its own value range. Q3 is still asleep waiting for a write latch to also perform cracking while Q1 blocks as well, as it needs a read lock for the aggregation but cannot proceed as Q2 is now cracking the column. When Q2 finishes with its crack select, both Q1 and Q2 acquire read latches and can now perform their aggregation operators in parallel. After this step, Q1 and Q2 are finished and Q3 may take a write latch and subsequently a read latch to perform its cracking and aggregation respectively.

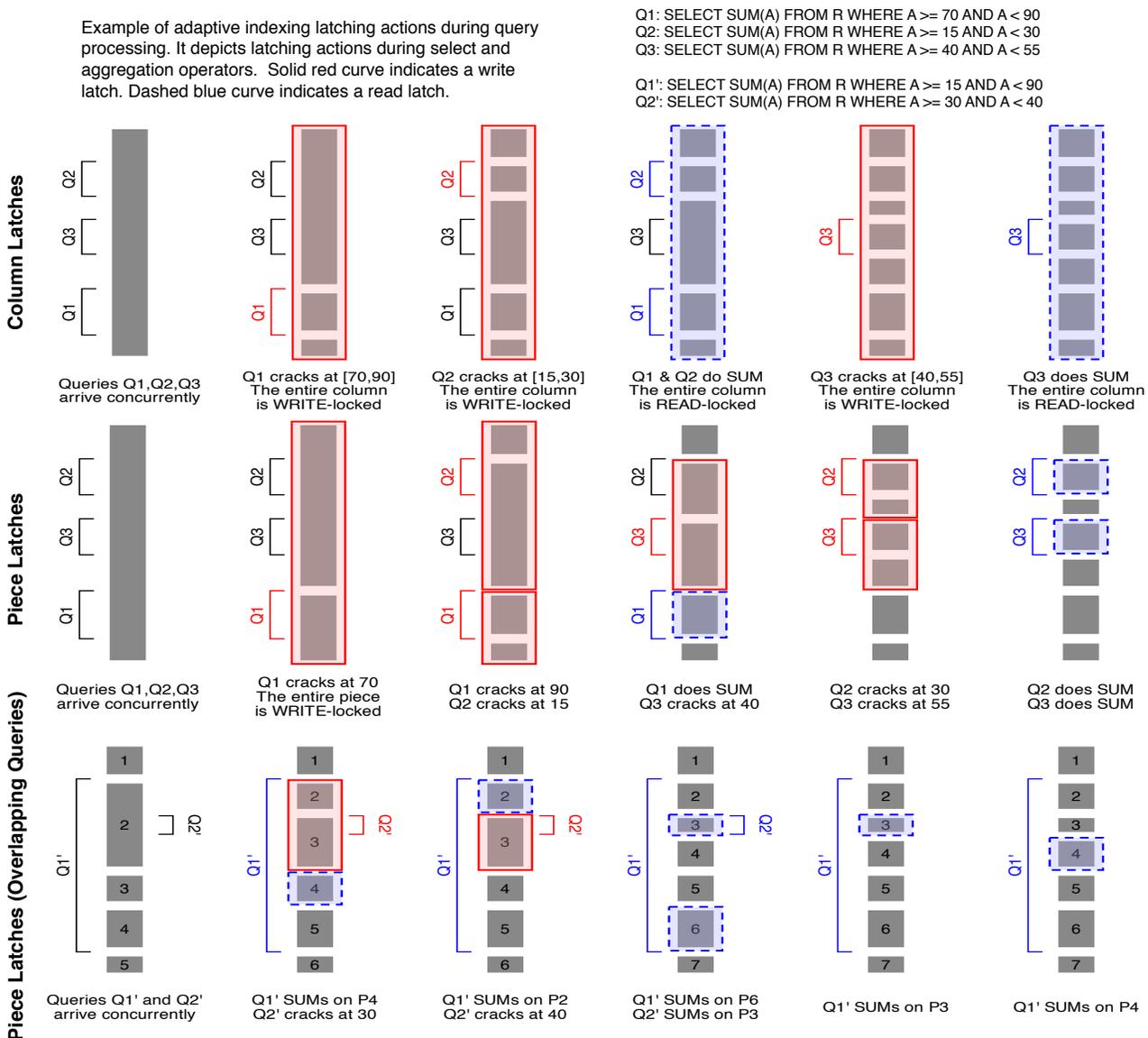

Figure 8: Concurrent queries with adaptive indexing.

**Piece-wise Latches.** As illustrated by the lower two-thirds in Figure 8, a natural enhancement is given by the fact that the index refinement of adaptive indexing in general and database cracking in particular only accesses a fraction of the index that has not yet been optimized for the requested key range. Hence, only the requested key range needs to be latched both in the cracker array and in the AVL tree. In fact, only the two pieces (segments) that contain the boundary values of the requested key range are physically reorganized. All pieces in between are fully covered by the requested key range, and thus not touched by the cracking select operator.

Figure 9 shows an example where a new query in an already cracked array, has to touch only two pieces; only the pieces where its low and high selection bound falls in. This results in a new array which is now cracked on the low and high bounds as well.

Hence, only the re-organization of the two boundary pieces needs to be protected by exclusive read-write latches, increasing the potential of concurrency even more. With piece-specific latches, two or more concurrent queries may proceed to crack the same column concurrently as long as they are cracking different pieces of the same column. Similarly, two or more queries may proceed to crack and perform aggregation on the same column concurrently, so long as they operate on different pieces; each distinct column piece can be accessed by one query at a time for cracking, while it can be accessed by multiple queries concurrently for aggregation.

**Optimizations.** An additional optimization is that the two cracks needed for each range select may be performed concurrently if they

662

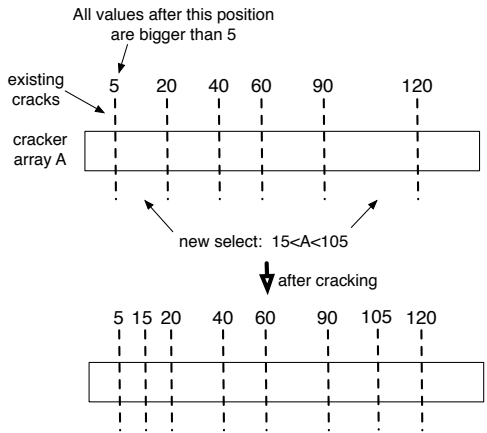

**Figure 9: Only need to touch two pieces during cracking.**

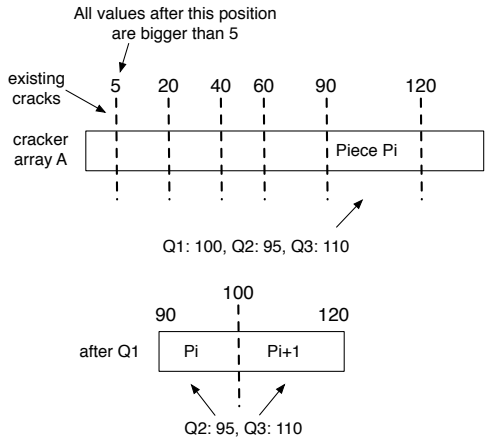

**Figure 10: Increasing concurrency with piece latching.**

are independent. For example, in Figure 9, the cracking action for bounds 15 and 105 can happen in parallel as they operate on different pieces. This way, even if there is a conflict for one of them the query actually proceeds with the second bound.

A crucial detail is that when two or more queries wait for a write latch over the same cracking piece, then upon waking up, the next query needs to re-determine its own bounds as the underlying piece has changed because of the previous query. The illustration in Figure 10 shows the various cases that may occur. Three queries need the same piece but only one can proceed. Once Q1 has finished, the structure of the underlying piece has changed, and Q2 and Q3 must reevaluate which area of the array they need to crack and where they need to latch. Every query achieves that by walking through the pieces of the array (the leaf nodes of the AVL-tree) starting from the original piece they tried to latch. For each piece they check whether their bound is in this range and if yes they try to latch this piece; otherwise they go on to the next. In Figure 10, Q2 still falls inside the original piece while Q3 is on the next one. In addition, given the creation of new pieces, now Q2 and Q3 may run in parallel as they no longer conflict.

Another optimization has to do with scheduling waiting queries in order to increase the concurrency potential. For example assume a piece with bounds on [0-100] and 5 waiting queries that want to crack on bounds Q1:20, Q2:30, Q3:50, Q4:70, Q5:90. The worst case scenario is if they wake up in the order of their requested bounds; e.g., Q1 wakes up first, then Q2, then Q3, etc. This scenario has the lowest potential for concurrency because the remaining queries must always wait. However, if Q3 runs first, the domain is split in half and the remaining queries may run in parallel. Our implementation uses a queue for each waiting query list in a given piece and will insert in the queue the queries with an insertion sort on their bounds. Then once the currently running query finishes, the next one will be the one which is in the middle of the queue.

**Example of Piece-wise Latching.** The middle third of Figure 8 illustrates piece-wise latching using exactly the same queries as the top part of this figure, which illustrates column-latching. As before, Q1 initializes and latches the entire, as-yet-uncracked, column. However, once Q1 has completed the cracking for its low bound, Q2 may proceed to start cracking for its own low bound while Q1 is cracking for its high bound concurrently. This is possible as after the first crack on the low bound of Q1, two independent pieces have been created. Subsequently, while Q1 is computing its aggregation with a read latch on its qualifying piece, the rest of the queries keep cracking the other pieces of the column.

The bottom third of Figure 8 depicts one more example of piece latching, where the requested ranges may vary across the incoming queries. With piece latching, cracking and aggregation queries may work concurrently so long as each cracking query has exclusive access to the piece being cracked. Two queries may crack different pieces concurrently, and two queries may perform aggregations in parallel in the same piece.

**Continuously Reduced Conflicts.** As the piece-wise discussion indicates, e.g., Figure 9, the pieces on the cracker array become smaller as the workload progresses. This is the very reason why adaptive indexing enjoys improved performance as we process more and more queries upon a given column. As the pieces of the index become smaller, we achieve both better filtering and also finer-granularity of latching. Together, these factors make the task of refining the index increasingly less expensive. Regarding concurrency conflicts this means that the period of time for which a query needs to hold the write latches decreases over time, which in turn allows more queries to run in parallel. In this way, the concurrency potential improves in an adaptive way; the more important a column is for the workload, the more chances appear to exploit concurrency as the workload evolves.

**Other Adaptive Indexing Methods.** The techniques presented here apply as is to the rest of the column-store designs for adaptive indexing which we introduced in [23]. This is because the ideas in [23] maintain the same underlying philosophy and follow the same column-store model. In addition, in future work we discuss interesting opportunities on how the status of the system during concurrency control may trigger new algorithm designs to improve performance even more, mainly by allowing for dynamic strategies which are driven by concurrency conflicts.

## 6. EXPERIMENTAL ANALYSIS

In this section, we report on a first implementation of concurrency control in adaptive indexing. The space of research is very broad, as we described in the previous sections. Here, we concentrate on the case of testing a column-store implementation of adaptive indexing using a full existing implementation of database cracking in the MonetDB open-source column-store.

**Set-up.** The set-up in the following experiments is as follows. We use a table of 100 million tuples populated with unique randomly distributed integers. The crucial part of adaptive indexing concerning concurrency is the index refinement as side effect of the selection over a base table. Consequently, to focus on this, we use simple range queries of the form.

Q1: select count(*) from R where $v_1 < A_1 < v_2$
Q2: select sum(A) from R where $v_1 < A_1 < v_2$

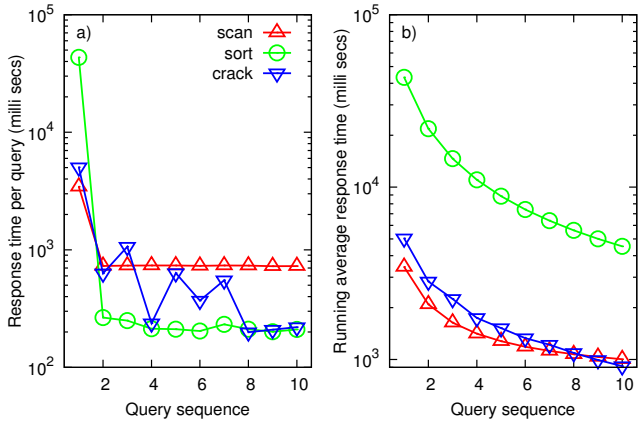
**Figure 11: Basic performance for sequential execution.**

The important difference between the two query types is that the second one has to do more work, i.e., both aggregation and selection/cracking.

In order to gauge the impact of concurrency on performance, we increase the number of clients submitting queries concurrently. We use lightweight queries in order to make the overhead of concurrency more prominent. The effect of more complex queries on adaptive indexing, e.g., TPC-H, can be seen in [22].

We use a 3.4 GHz Intel Core i7-2600 quad-core CPU equipped with 32 KB L1 cache and 256 KB L2 cache per core, 8 MB shared L3 cache and 16 GB RAM. The operating system is Fedora 14.

## 6.1 Basic Performance

This first experiment establishes context by illustrating the basic trade-offs of adaptive indexing as distinct from any concurrency related overhead. The scenario is a completely dynamic environment. We assume no workload knowledge and idle time to prepare the system. The only given is that the data is assumed loaded in its basic form. Immediately after the data is loaded, queries begin to arrive in a steady stream with no "think-time."

The experiment compares three approaches using queries of type Q1. In the default case, the system accesses the data using plain scans, with no indexing mechanism present. At the other extreme, we consider the case of a very active approach that resembles a traditional indexing mechanism: when the first query arrives, we build the complete index before we evaluate the query, which can then exploit this index. The benefit is then available to all future relevant queries. In our implementation over a column-store it is sufficient to completely sort the relevant column(s) and then use binary search to access them.

We use adaptive indexing via a complete implementation of database cracking over MonetDB. Query processing operators reorganize relevant columns and tree structures on-the-fly to reflect the knowledge gained by each query. All changes happen automatically as part of query processing and not as an afterthought.

Figure 11(a) compares the basic performance of these three approaches in terms of per-query response time for running 10 queries serially one after the other through a single database client. The queries use random range predicates with a stable 10% selectivity. The default scan-based approach has a rather stable behavior. The first query is slightly slower, fetching the data from disk. The full indexing approach, labeled "sort" in the figure, shows a significant overhead when building the index with the first query; then enjoys great performance from the second query onwards.

The problem with the scan approach is that it does not exploit past knowledge, resulting in relatively slow performance through-out the span of a workload. The problem with the full indexing approach is that it significantly penalizes the very first query. If this query were an outlier, or if the workload span turned out to consist of only a few queries, then this extra overhead may never pay off. Figure 11(b) visualizes this by depicting the running average response time for the same experiment. 10 queries are far from enough to amortize the high investment of building the full index with (or before) the first query.

Adaptive indexing solves the above problems in dynamic environments. Figure 11(a) shows how it maintains a lightweight first touch to the workload but at the same time, it continuously learns and improves performance in a seamless way, without overpenalizing queries. Performance improves continually and almost immediately in response to the workload. The more queries arrive, the more performance improves. Figure 11(b) confirms that the low initial investment pays back quickly; after only 8 queries the initial investment has paid off and the average per-query response time of adaptive indexing becomes less than that of a basic scan approach.

The performance seen in this experiment is representative of the adaptive indexing behavior. The interested reader can refer to previous papers for in-depth analyses regarding multiple parameters, e.g., skew, updates, multi-column indexes, etc. [20, 21, 22, 11, 12, 14, 23]. In the rest of the following analysis, we focus solely on concurrency control issues.

## 6.2 Concurrency Control

Let us now focus on how concurrency control impacts performance. For ease of presentation, in this section we first present a broad analysis. Then, the next section will dive deeper into analyzing the behavior for various parameters and it also presents piece latches in more detail.

The set-up of our next experiment is as follows. We repeatedly run a sequence of 1024 random range queries of 0.01% selectivity and of type Q2. Each time, we increase the number of concurrent streams. Selectivity is purposely kept high to more clearly isolate the costs of the select operator, i.e., do not let aggregation operators hide any overheads. The next section studies this parameter in more depth. In more detail, we run the serial case where one client runs all 1024 queries, one after the other. Then, we use 2 clients that start at the same time and each one fires 512 queries. Then, we repeat the experiment by starting 4 clients at the same time and each one fires 256 queries, and so on. We go up to the limit of 32 clients which is the threshold that our experimentation platform, MonetDB, puts in order to throttle the incoming clients and control the amount of concurrent query threads. For every run we use exactly the same queries and in the same order.

Figures 12(a) and (b) depict the results for plain scan, full indexing (sort) and for database cracking, using piece latches. In both figures, the $x$-axis lists the increasing number of concurrent clients. In Figure 12(a), the $y$-axis represents the total elapsed time needed to run all 1024 queries. In Figure 12(b), the $y$-axis shows the "inverted" results for the same experiment, i.e., depicting throughput in terms of queries per second rather than total execution time for all 1024 queries.

For all approaches we see a rather similar trend, i.e., performance shows a linear decrease in total execution time and consequently a linear increase in throughput when going from one over two to four clients, i.e., up to the number of CPU cores in our system. Then performance peaks at 8 clients, before leveling out and remaining quite stable up to the case of 32 clients running 32 queries each.

The relative behavior between the three different approaches remains the same, regardless the number of concurrent queries. Scan suffers from having to scan the complete column with each query.

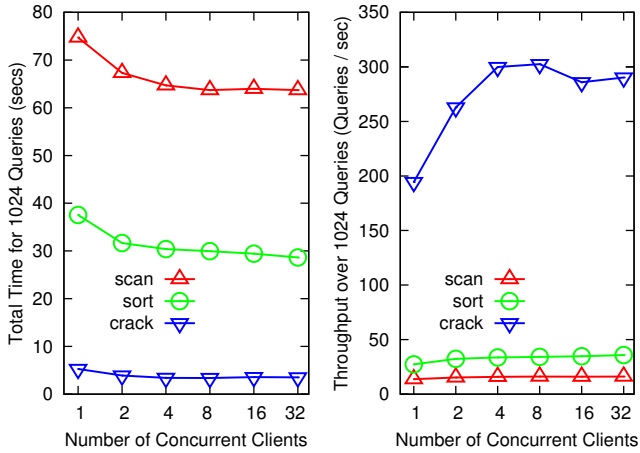

Figure 12: Effect of Concurrency Control on Total Time.

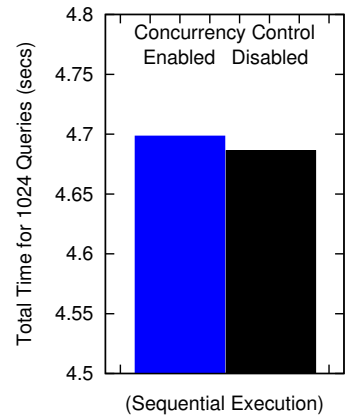

Figure 13: Concurrency Control Overhead of Adaptive Indexing (with and without concurrency control.)

Full indexing improves over plain scans, but suffers from having to build the complete index via sorting the column. On the other hand, adaptive indexing maintains its competitive advantage and adaptive behavior even with concurrent queries.

We point out that due to their purely read-only data access, neither scans nor binary search actions used in full indexing require any concurrency control during the actual query processing. Adaptive indexing on the other hand has to incur concurrency control costs as it turns read queries into write queries. Nevertheless, its performance remains unaffected.

All in all, the results in Figure 12 confirm that although adaptive indexing introduces write access for conceptually read-only queries, concurrency is not only possible but also beneficial. Instead of having issues with multiple queries touching the same data, adaptive indexing manages to parallelize queries and benefit from that. The amount of index refinement — and hence the length of the critical part of the query — becomes less and less with every query, quickly vanishing behind the non-critical parts that can be executed in parallel. The next section discusses these issues in more detail.

Figure 13 provides insight on the overhead of concurrency in adaptive indexing. Using the same set-up as before, we run the 1024 queries using a single client. This way all queries run sequentially one after the other. Hence, no concurrency control is required to protect data structures and ensure correct execution. We repeat the experiment twice. For the first run, the concurrency control mechanisms are enabled but for the second run we disable all concurrency control activities. Given sequential execution, the sole difference between the two runs is that the first performs concurrency control related actions (mainly managing, acquiring and releasing latches), while the second one does not. Thus, the difference in execution time between both runs is the administrative overhead required for the concurrency control mechanisms of adaptive indexing. Figure 13 reports the total costs to run all 1024 queries. For the complete sequence of 1024 queries, the concurrency control overhead is less than 1%.

### 6.3 Detailed Analysis

Having seen a generic analysis in the previous section, we now go into more detail to explain the behavior seen under various parameters. Figure 14 depicts the results for our next experiment. We use the same set-up as before, i.e., 1024 random queries and a varying number of clients ranging from 1 (sequential execution) to 32 clients. Here, we also present the behavior of piece vs column latches as well as we study both queries of type Q1 and of type Q2. In addition, we run the experiment for various selectivity factors for each case; queries remain random but selectivity varies. The graphs in Figure 14 depict the total time needed to run all queries in each case, i.e., the time reported is the time perceived by the last client to receive all answers for all its queries.

Figures 14(a) and (b) demonstrate the performance for queries of type Q1 with column and piece latches respectively. Excluding the low selectivity case (90%), performance is rather similar for all selectivity runs. This is true both for column and piece latches. With selectivity 90% all queries use low and high bounds in their range selection predicates that are focused on only 10% of the column. As a result adaptive indexing improves even faster by refining these areas of the column faster compared to other selectivity cases.

When comparing column and piece latches in Figures 14(a) and (b) we see that piece latches bring significantly more improvements to the adaptive indexing performance. With column latches performance is rather stable which means that adaptive indexing is not affected by concurrent queries but at the same time it does not manage to exploit opportunities for parallelism. This effect is even more noticeable in Figures 14(c) and (d) where we study queries of type Q2. For such queries an aggregation on the selection column needs to be performed. For this reason, an aggregation operator needs to hold a read latch while going through all qualifying tuples, computing the aggregation. During this time, no cracking can happen and thus no other select operator may run. Only read latches are allowed, e.g., for other aggregation operators of other queries. In the case of column latches, this results in a significant penalty; the whole column needs to be latched.

The lower the selectivity, the higher this penalty as the time needed to perform the aggregation increases (due to more tuples qualifying the selections) and dominates the total query cost. On the other hand, with piece latches we allow many queries to run in parallel multiple kinds of previously conflicting operations over the same column as long as they operate on different pieces. Now two queries may crack in parallel two or more different pieces or may crack in one piece and run aggregation on others. This increased parallelism allows piece latching to materialize an even more significant benefit which in the case of Q2 type queries becomes more evident due to the need to maintain read latches for a longer period of time. In this way, this phenomenon becomes more apparent as the selectivity decreases in Figure 14(d).

Figure 15 gives more insight in the above results by breaking down the time of individual queries. It depicts the wait time and the crack time for each individual query as the workload sequence evolves. This is for the case of queries of type Q2 using piece

665

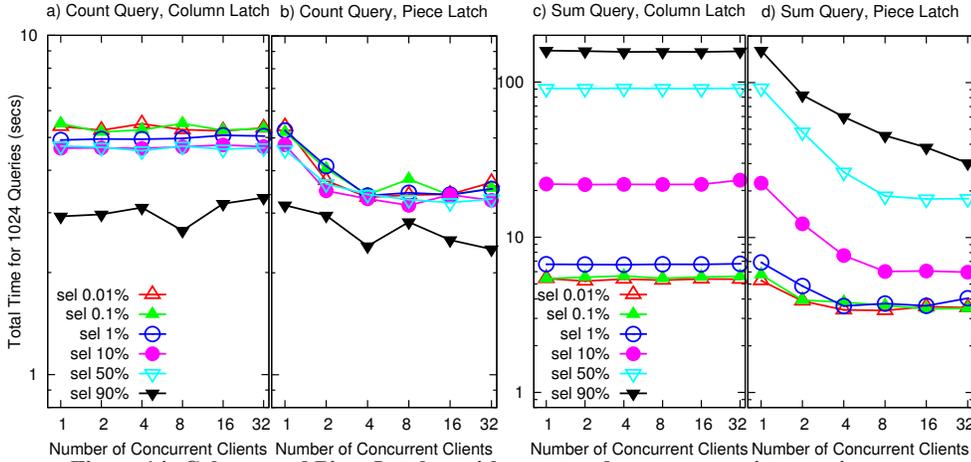
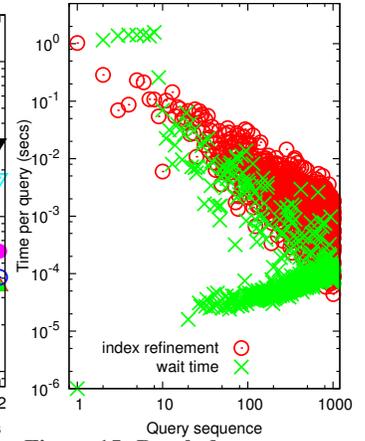

Figure 14: Column and Piece Latches with count and sum aggregation queries.

Figure 15: Break down costs.

latches with 50% selectivity and with 8 clients. The wait time is defined as the time each query spends in waiting to acquire a latch. For each query, the number plotted reflects all waiting time, i.e., both for write latches during the crack select operator as well as the waiting time for read latches during the aggregation operator. In addition, the time reflects the time needed to acquire all latches for all relevant pieces in each operator. The crack time is defined as the time spent purely on refining the index during the select operator (under write latches).

Figure 15 shows that the crack costs follow the behavior which was observed in past adaptive indexing papers as well, i.e., the more we touch a specific column, the more the index is refined. As pieces become smaller due to more fine grained indexing, subsequent index refinement operations become faster. This is what brings the adaptive behavior and Figure 15 shows that adaptive indexing maintains this behavior even during concurrent queries.

The second observation from Figure 15 is that the waiting time, i.e., the concurrency control conflicts, shows a similar behavior; it decreases as the workload sequence evolves. Naturally, the very first query does not have to wait at all, depicting a zero cost waiting time in Figure 15. The next 7 queries though have to wait until the first one finishes cracking the column. This is 7 queries because we use 8 concurrent clients in this experiment and they all have to wait because when the experiment starts there is no cracking index, meaning that the first query has to latch the complete column. Once the first query adds some partitioning, then the concurrency opportunities increase and soon after a few queries have cracked the column, the waiting times decrease.

The main bottleneck in the crack select where the write latches are required is the index refinement time. As this time decreases in Figure 15, the concurrency conflicts decrease as well. A closer observation on the waiting time in Figure 15 shows that the wait time almost matches the crack time behavior. For some queries (including the first one) the wait time is minimal as they happen to arrive at a time that the needed piece is free of latches. For the rest of the queries, the wait time follows a continuously decreasing trend similar to crack time; the crack time of one query is in practice the wait time for another query, waiting for a given column piece.

Thus, by using short latching periods and quickly releasing latches as soon as possible, adaptive indexing manages to exploit concurrent queries as opposed to suffering from them. In addition, it is interesting to notice that since adaptive indexing gains continuously more and more knowledge about the data, these latching periods become ever shorter which improves performance even more.

## 7. FUTURE WORK

In this section, we discuss on-going efforts and future work with regard to concurrency control in adaptive indexing.

Each distinct adaptive indexing algorithm follows its own strategy on how the index is incrementally and adaptively refined. For example, database cracking follows a very "lazy" approach with only small index refinement steps [20]. Adaptive merging on the other hand introduces sorting steps to reach optimal performance faster [14]. The hybrid adaptive indexing algorithms presented in [23] use a collection of steps such as radix clustering, plain cracking and sorting steps. In any case though, the decision on which algorithm to use is fixed a priori.

We observe that with concurrency control in mind, we have the opportunity to optimize the system performance and adaptivity with concurrent queries by revisiting the choice for strictly deciding adaptive indexing algorithms a priori.

*"Active" Algorithms.* In adaptive indexing, read-only queries may trigger physical structural changes, which in turn can introduce read-write conflicts. As long as these incremental refinement steps continue, we may encounter concurrency issues. However, once an index has reached an optimal state for the current workload, queries cease to trigger refinement actions and any accompanying concurrency issues. From this, one might conclude that "active" strategies that aggressively refine the index could be less prone to concurrency issues because there would be reduced opportunity for conflict to occur at the cost of incurring a potentially high cost during the active steps.

*"Lazy" Algorithms.* The counterpart of active strategies develops "lazy" techniques where certain queries refrain from introducing any side effects and thus immediately avoid or significantly reduce the need for concurrency control. For example, we can deliberately ensure that only one query at a time operates with side effects on a given part of the data. This would reduce write contention and thus enable higher concurrency levels, but at the cost of slower rates of refinement.

*Dynamic Algorithms.* Until now, adaptive refinement actions have *reacted* to each query independently, deciding for each individual query how to refine data structures for the current workload. One natural next step is thus to consider using groups of multiple queries to guide index refinement. For example, all queries waiting for a given index piece may contribute to a more high level strategy on how to refine the index for the given value range. For example, one idea is to invest in algorithms that in one step refine the index for multiple query requests. The key here is that we both exploit the fact that we know all the index refinement requests and to increase concurrency at the same time. For this reason, we envision that the strict and fixed policies of which adaptive indexing algorithm is used have to revisited. Depending on the status the system might benefit from using different adaptive indexing algorithms at different times, depending on the number of queries, status, etc. Even in the same column, we could use different adaptive indexing algorithms for different parts of the domain if the access patterns dictate doing so.

*Multiple Indexes.* Finally, our prior work has focused on the case that each query refines one or more indexes independently. Other models are equally possible, raising a plethora of interesting issues and opportunities both for query processing and for concurrency control, e.g., adaptive indexing mechanisms could leverage create and refine multiple index structures in a unified way as a side effect of processing a single query.

# 8. SUMMARY AND CONCLUSIONS

Recent papers have introduced adaptive indexing in the forms of database cracking and adaptive merging. The main idea shared by both techniques is on-demand index construction and optimization as side effects of query execution. At first glance, this seems to turn read-only queries into update transactions, triggering the question whether the anticipated concurrency control overhead will prohibit the use of adaptive indexing in multi-user scenarios. In this paper, we address this question and show that with judicious application and extension of prior work, concurrency control conflicts and overheads can be reduced to practical or even negligible levels.

The key observation is that adaptive indexing applies only *structural* modifications to the physical representation of the index but leaves the logical *contents* of the index unmodified. This relaxes the constraints and requirements during adaptive indexing compared to those considered for traditional index updates. Furthermore, we observe that even those structural changes are optional. Using adaptive merging and database cracking as examples, we introduce concrete implementations of our new techniques. The experimental evaluation of our implementation of concurrency control for database cracking demonstrates that the performance overhead of concurrency control during structural updates is minimal, and that adaptive early termination alleviates problems with concurrency control in adaptive indexes.